**Shock-Induced Flows through Packed Beds: Transient Regimes**

**Yuri M. Shtemler[1], Isaac R. Shreiber, Alex Britan**

Department of Mechanical Engineering,

Ben-Gurion University of the Negev,

P.O.B. 653, Beer-Sheva 84105, Israel

[1]shtemler@bgumail.bgu.ac.il

Short title: Shock-Induced Flows through Packed Beds

**Abstract** The early stage of the transient regimes in the shock-induced flows within solid-packed beds are investigated in the linear longwave and high-frequency approximation. The transient resistance law is refined as the Duhamel-time integral that follows from the general concept of dynamic tortuosity and compressibility of the packed beds. A closed-form solution is expected to describe accurately the early stage of the transient regime flow and is in qualitative agreement with available experimental data.

## 1. Introduction

In the present study a pressure wave filtrating through a gas-filled packed bed induced by a shock wave incident on a packed-bed sample is considered (Fig. 1). The moving wave involves the fluid behind it into the motion. As a result, unsteady boundary layers arise at the packed-bed capillary walls, produce unsteady resistance and move from the capillary walls to the pore centers. A simple illustration of a shock-induced flow in a packed-bed is shown schematically in Fig. 2 within a single packed-bed capillary of the equivalent hydraulic radius. The conventional models consider the hydraulically resisted flows after the unsteady boundary layers reach the centers of the capillaries. Then the resulting flow tends asymptotically to the steady-state Hagen-Poiseuille flow that utilizes the Darcy resistance law. This approach often proves to be too rough, markedly underestimating the impact of hydraulic resistance in the early stage of the boundary-layer formation as it is demonstrated by experiments with combustion-driven flows through packed beds (e.g. Frolov and Gelfand 1991). In the combustion systems filtration flows arise due to the flame front moving through packed beds, and in this respect they are similar to shock-induced filtration. To fit the modeling results for shock- or combustion-driven filtration with experimental data various empirical correlations (Ergun, Forchheimer etc.) for resistance laws are employed (see e.g. Rogg *et al*. 1985, Smeulders *et al*. 1992, Ben-Dor *et al*. 1997, Britan *et al*. 1997, Brailovsky *et al*. 1997, Brailovsky and Sivashinsky 2000 *a,b*). In this connection note that for numerous hydrodynamic systems the transient resistance law may be captured by utilization of the Duhamel-time integrals (e.g. for flows within tubes with deformable walls, flows of bubbly liquids and gas-liquid foams, filtration within packed beds with skeleton memory or micro-heterogeneity, see Holmboe and Rouleau 1967; Nakoryakov *et al*. 1993; Goldfarb *et al*. 1994, Hanyga and Rok 2000; Fellah and Depollier 2000; Bednarik and Konichec 2002 etc.). The Duhamel-time resistance law is in compliance with the concept of dynamic tortuosity of the packed beds (Johnson *et al*. 1987), which has been generalized to take into account thermal effects (the general concept of dynamic tortuosity and compressibility developed by Champoux and Allard 1991, Lafarge *et al*. 1997, Allard *et al*. 1998). To derive the generalized resistance law the viscous and thermal boundary layers are assumed to be of the small thicknesses $\delta$ and $\delta'$ compared with the pore sizes appropriate for the capillary friction and heat exchange ($\delta' \sim \delta$, $\Lambda' \sim \Lambda$, Fig. 2).

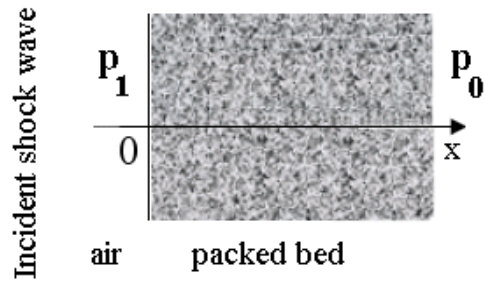

**Figure 1.** Shock-induced flow in a fluid-saturated packed bed (schematic).

$\Delta p = p_1 - p_0$ is the pressure head through the packed bed ($p_0$ is the atmospheric pressure).

The paper is organized as follows. Section 2 contains the closed form solution for the longwave and high-frequency approximation of the transient regime flow through packed beds, and transition to the non-dimensional variables (Section 2.1); comparison of the modeling results with experimental data, and verification of the main qualifications adopted in the present study (Section 2.2). The summary and discussion are given in Section 3. The basic relations governing the transient macroscopic pressure and velocity fields in the high frequency approximation are presented in Appendix A.

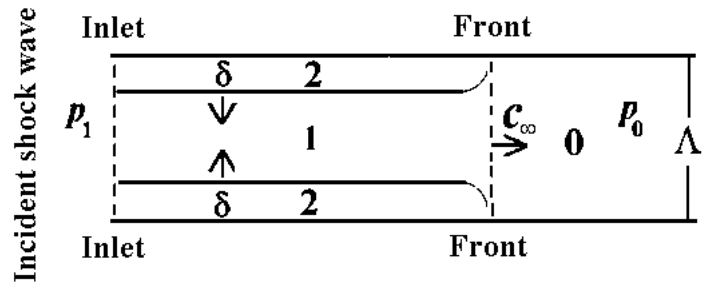

**Figure 2.** Scheme of the transient regime of shock-induced flows in a packed bed capillary

(arrows indicate the moving directions).

$\Delta p = p_1 - p_0$ is the pressure head through the packed bed ( $p_0$ is the atmospheric pressure); $c_\infty$ is the

wave-front velocity; dash lines denote the inlet to the packed bed sample and positions of the wave front;

**0** is the packed-bed zone with the fluid at the rest state; **1** is the transient zone; **2** are the viscous and thermal boundary layers of thickness , $\delta' \sim \delta$ ; $\Lambda$ and $\Lambda'$ are the viscous and thermal characteristic pore sizes

( $\Lambda' \sim \Lambda$ ).

## 2. Mathematical Modeling of Transient Regimes in Shock-Induced Flows

Assuming the system is at rest at $t < 0$, let us consider the fluid (air) -saturated packed bed located at $x > 0$,

and describe the pressure wave filtration through the packed bed due to the arrival of a shock wave incident to the edge of the packed-bed sample from $x < 0$ at $t = 0$. The system is characterized by the total pressure head

$\Delta p \equiv const$ through the packed-bed sample. Neglecting the packed bed elasticity, the linear longwave and

high-frequency approximation is adopted.

*2.1 Closed-form solution in the high-frequency approximation for the transient regime flow in a packed bed*

The governing macroscopic equations of flow in the fluid-saturated packed bed that occupies the half-space $x > 0$ are as follows (see Appendix):

$$\frac{\partial \overline{\rho}}{\partial t} + \rho_0 \frac{\partial \overline{u}}{\partial x} = 0, \ \rho_0 \alpha_\infty \frac{\partial \overline{u}}{\partial t} + \frac{\partial \overline{p}}{\partial x} = F, \ \overline{p} + G = c_0^2 \overline{\rho}, \tag{1}$$

$$F = \frac{2\sqrt{\nu}}{\Lambda} \frac{\partial^{1/2} \overline{p}}{\partial x \partial t^{-1/2}}, \qquad G = \frac{2(\gamma-1)\sqrt{\nu'}}{\Lambda'} \frac{\partial^{-1/2} \overline{p}}{\partial t^{-1/2}}, \qquad \alpha_\infty = 1.37.$$

Here the hydraulic resistance force, $F$, the effective thermal resistance, $G$, and the tortuosity value, $\alpha_\infty$, are given in the high-frequency limit; the problem is linearized near the main rest state that is characterized by the total pressure and density at the normal atmospheric conditions, $p_0$ and $\rho_0$, and the fluid velocity $u_0 \equiv 0$; $c_0 = \sqrt{\gamma p_0 / \rho_0}$ is the adiabatic sound velocity, $\gamma$ is the adiabatic constant; $x$ and $t$ are the longitudinal coordinate and time; the macroscopic values of the longitudinal velocity, pressure and density, $\overline{u}, \overline{p}$ and $\overline{\rho}$ are obtained by averaging corresponding microscopic values; $\nu$ and $\nu'$ are the kinematic viscosity and thermal diffusivity of the fluid; $\Lambda$ and $\Lambda'$ are the viscous and thermal characteristic pore sizes appropriate for the capillary friction and heat exchange. The boundary and initial conditions explained at the beginning of the current section are assumed as follows:

$$x = 0 : \overline{p} = \Delta p \cdot H(t), \quad t = 0 : \overline{p} = 0, \ \frac{\partial \overline{p}}{\partial t} = 0, \tag{2}$$

$H(t)$ is the Heaviside step function; $\Delta p = p_1 - p_0 \equiv const$ is the total pressure head through the packed bed.

To properly choose the characteristic scales, let us consider the following system conventional for acoustic waves. Accounting for the packed-bed tortuosity, but neglecting viscous and thermal effects, Eqs. (1), (2) are reduced to:

$$\frac{\partial \overline{\rho}}{\partial t} + \rho_0 \frac{\partial \overline{u}}{\partial x} = 0, \quad \rho_0 \alpha_\infty \frac{\partial \overline{u}}{\partial t} + \frac{\partial \overline{p}}{\partial x} = 0, \quad \overline{p} = c_0^2 \overline{\rho}, \quad x = 0: \overline{p} = \Delta p \cdot H(t), \quad t = 0: \overline{p} = 0, \frac{\partial \overline{p}}{\partial t} = 0. \quad (3)$$

System (3) has the following wave solution in the half-space $x > 0$ occupied by the packed bed:

$$\overline{p} = \Delta p \cdot H(c_\infty t - x), \quad \overline{\rho} = \frac{\Delta p}{c_0^2} \cdot H(c_\infty t - x), \quad \overline{u} = \frac{\Delta p}{\rho_0 \alpha_\infty c_\infty} \cdot H(c_\infty t - x), \quad (4)$$

where $c_\infty = c_0 / \sqrt{\alpha_\infty}$ is the high-frequency limit of sound velocity. According to Eqs. (4) and additionally setting $t^* = \Lambda^2 /(4\nu)$, a natural scaling in this problem is

$$p^* = |\Delta p|, \frac{\rho^*}{\rho_0} = \frac{p^*}{\rho_0 c_0^2}, \quad u^* = c_\infty \frac{\rho^*}{\rho_0}, \quad t^* = \frac{\Lambda^2}{4\nu}, \quad x^* \equiv c_\infty t^*. \quad (5)$$

To present the above relations in dimensionless form the physical variables were transformed into nondimensional variables $\hat{f} = f / f^*$, where $f$ and $f^*$ stand for any of the dimensional variables and their characteristic scales. Equations (1)-(2), made dimensionless and omitting with no confusion the hats everywhere below, if not said otherwise, read:

$$\frac{\partial \overline{\rho}}{\partial t} + \frac{\partial \overline{u}}{\partial x} = 0, \quad \frac{\partial \overline{u}}{\partial t} + \frac{\partial \overline{p}}{\partial x} = F, \quad \overline{p} + G = \overline{\rho} \quad , \quad x = 0: \overline{p} = H(t), \quad t = 0: \overline{p} = 0, \frac{\partial \overline{p}}{\partial t} = 0. \quad (6)$$

Here the dimensionles hydraulic resistance force and the effective thermal resistance are:

$$F = \frac{\partial^{1/2} \overline{p}}{\partial x \partial t^{-1/2}}, \quad G = (\gamma - 1)\frac{\Lambda}{\Lambda'}\frac{\partial^{1/2} \overline{p}}{\partial x \partial t^{-1/2}}.$$

Substituting the explicit expressions for $F$ and $G$ within the accuracy adopted reduces problem (6) to:

$$\left\{ 1 + \left[ 1 + (\gamma - 1)\frac{\Lambda}{\Lambda'}\sqrt{\frac{\nu'}{\nu}} \right]\frac{\partial^{-1/2}}{\partial t^{-1/2}} \right\} \frac{\partial^2 \overline{p}}{\partial t^2} - \frac{\partial^2 \overline{p}}{\partial x^2} = 0, \quad x = 0: \overline{p} = H(t), \quad t = 0: \overline{p} = 0, \frac{\partial \overline{p}}{\partial t} = 0. \quad (7)$$

The fractional derivative of the order of $3/2$ along with conventional second order derivatives distinguishes the resulting equation for pressure from the diffusion and wave equations describing diffusion waves (Barenblatt *et al.* 1990) and conventional acoustic waves (Eqs. (3)). By taking the square root of the operator in problem (7) it can be reduced to the leading-order:

$$\frac{\partial \bar{p}}{\partial x} + \left\{ 1 + \frac{1}{2}\left[ 1 + (\gamma - 1)\frac{\Lambda}{\Lambda'}\sqrt{\frac{v'}{v}}\right]\frac{\partial^{-1/2}}{\partial t^{-1/2}}\right\}\frac{\partial \bar{p}}{\partial t} = 0, \quad x = 0: \bar{p} = H(t), \quad t = 0: \bar{p} = 0 \,. \tag{8}$$

Application of the Laplace transform to problem (8) yields the following solution:

$$\bar{p} = H(t - x)\text{erfc}\left\{ \frac{1}{4}\left[ 1 + (\gamma - 1)\frac{\Lambda}{\Lambda'}\sqrt{\frac{v'}{v}}\right]\frac{x}{\sqrt{t - x}}\right\}, \tag{9}$$

where the first factor describes the pressure wave propagation without attenuation, while the second describes its irreversible viscous and thermal spreading.

In some cases the problem may be additionally simplified. Thus, for air-saturated random packed beds the parameters of the problem may be estimated as follows (Allard *et al.* 1998):

$$\gamma = 1.4, \ \text{Pr} = \frac{v}{v'} \approx 0.7, \ \frac{\Lambda}{\Lambda'} \approx 0.28, \ \alpha_\infty = 1.37 \,. \tag{10}$$

Here $\text{Pr}$ is the Prandtl number. Using (10) for estimation of the coefficient in Eq. (9), one has:

$$(\gamma - 1)\frac{\Lambda}{\Lambda'}\sqrt{\frac{v'}{v}} \approx 0.135 \,. \tag{11}$$

In neglect of this value compared with unit, i.e. in neglect of the input of the thermal boundary layer compared with the input of the viscous boundary layer, Eq. (9) yields

$$\bar{p} = H(t - x)\text{erfc}\left\{ \frac{1}{4}\frac{x}{\sqrt{t - x}}\right\} \,. \tag{12}$$

*2.2. Comparison of the modeling results and experimental data*

In most shock tube experiments filtration phenomena in the packed beds are frequently investigated based on the pressure measurements (Britan *et al*. 1997; Gelfand *et al*. 1986). Moreover, to register the time interval for the wave front to traverse the known "base distance" between the several pressure transducers (this method is usually referred to as the "base distance" method) an average velocity of the wave-front leading point can be measured. Although for pressure waves induced in packed beds by weak shocks and rarefaction waves these measurements should result in the constant value of sound velocity in gas phase, experiments (Britan *et al*. 1997; Gelfand *et al*. 1986) demonstrate an unexpected dependence of the wave-front velocity on the sizes of the packed-bed particles. In fact, for the nondeformable packed-bed particles the wave-propagation velocity should be equal to the gas-sound velocity. This inconsistency can be explained by using the above-described dynamics of the transient pressure waves.

For the sake of convenient comparison with experimental data let us write out expressions for the dimensional velocity, density and pressure using Eq. (12):

$$M = \frac{\bar{u}}{c_\infty} = \frac{\bar{\rho}}{\rho_0} = \frac{\bar{p}}{p_0} = \frac{\Delta p}{p_0} H(c_\infty t - x) E(\theta), \ \ E(\theta) = \text{erfc}(\theta), \ \theta(x,t) = \frac{x}{2\Lambda} \sqrt{\frac{\nu}{c_\infty} \frac{1}{t c_\infty - x}}, \tag{13}$$

where fields are presented in the half-space $x > 0$ occupied by the packed bed, $E$ is the decrement factor, $M$ is the local Mach number.

To relate the parameters of the flow within a packed bed capillary to the parameters of the corresponding packed bed, let us assume that a random packed bed of equal-size spherical particles is described as a bundle of non-intersecting capillaries. Then the dependence of the hydraulic radius $\Lambda$ on the porosity $\varphi$ and radius $\Lambda_p$ of the packed bed particles may be approximated as follows (Ng 1986): $\Lambda = \Lambda_p \sqrt{2/3\varphi/(1-\varphi)}$. Simulated pressure profiles vs time are presented in Fig. 3$a$-3$c$ by solid lines for different locations of pressure sensors. As should be expected for the nondeformable packed-bed particles, the wave propagation velocity (defined as the velocity of the front leading point of the pressure wave) equals the gas-sound velocity independent of particles sizes. The pressure profiles demonstrate that the wave propagation in packed beds with large particle sizes is similar to step-wise pressure profiles within wide tubes, while for small particles the pressure waves strongly attenuate and spread. In particular, the near-zero flat parts of pressure profiles are of near-zero length at large particle sizes and of increasing length with decreasing particle sizes. The results of simulations are in qualitative agreement with experimental data in some respects: strong dependence of pressure profiles on particle sizes; apparent dependence of the wave propagation velocity on the particle sizes.

As it evidently follows from Figs. 3 and 4, the time duration $\Delta t$ which it takes for the wave front to pass the known "base distance" between two fixed positions is constant regardless of the size of the packed bed particles and initial pressure values $p_0$. In practice, however, the reading accuracy of the pressure trace origin (where $\hat{p}=0$) depends on dispersion of the wave front and on uncertainty in the pressure measurements, $\hat{p}_*$, shown by horizontal dashed lines in Fig. 3. In fact, when the particle radius is large (see Fig. 3$a$), the step-wise pressure profile is changed very slightly and the pressure trace origin is clearly specified for two base positions. For smaller particles (Figs. 3$b$ and 3$c$) distortion of the pressure front causes significant misleading in the origin of the simulated pressure trace at $x_2 = 0.45m$. Apparent increase in the time duration $\delta t$ is shown by the vertical dotted line and is more than twice the real characteristic $\Delta t$ of the phenomenon in Fig. 3$c$.

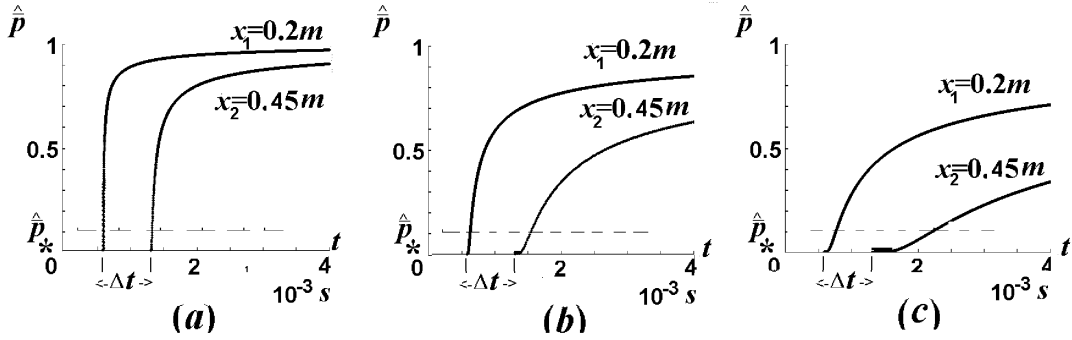

**Figure 3.** Modeled normalized pressure $\hat{p}$ and pressure-sensor sensitivity $\hat{p}_*$ vs time $t$

for 2 locations of pressure sensors and 3 values of the particle radius $\Lambda_p$ ($\varphi = 0.4$, air flow):

$(a)\,\Lambda_p = 2.5 \cdot 10^{-3}\,m$ , $(b)\,\Lambda_p = 0.75 \cdot 10^{-3}\,m$ , $(c)\,\Lambda_p = 0.35 \cdot 10^{-3}\,m$ .

$\hat{\bar{p}} = \bar{p}/\Delta p, \hat{\bar{p}}_* = \bar{p}_*/\Delta p,\, \Delta t = t_2 - t_1$; $x_2 - x_1$ is the "base distance";

$\Delta p = p_1 - p_0$ is the pressure head through the packed bed; $\Delta t$ is the time duration of the wave front passage

of the "base distance"; $\delta t$ is the apparent increase in the time duration.

Equation (13) illustrates that the wave leading point propagates with sound velocity $c_\infty$ regardless of the particle size, porosity, pressure values etc. In fact, non-accuracy of pressure measurements may significantly influence this result. Thus, standard accuracy of pressure measurements does not usually exceed 10% of the measured pressure-range. The time necessary for passing the distance between two points corresponding to $\hat{\bar{p}}_*$=0.075 is about 1.5 times less at $\Lambda_p = 0.75 \cdot 10^{-3}\,m$ (Fig. 3b) than at $\Lambda_p = 0.35 \cdot 10^{-3}\,m$ (Fig. 3c). Hence the wave "velocity" $c_\infty = (x_2 - x_1)/\Delta t$ should be about 1.5 times higher at $\Lambda_p = 0.75 \cdot 10^{-3}\,m$ than at $\Lambda_p = 0.35 \cdot 10^{-3}\,m$ $((x_2 - x_1) \approx 0.25m$, Figs. 3 b,c). The strong apparent dependence of the velocity of the wave-front leading point on the particle sizes is illustrated by Fig. 4. This is in fair qualitative agreement with the measurement results, where such dependence obtained by the "base distance" method has been noted (Ben-Dor *et al.* 1997, $p_1 \approx 2.4\ bar$, $p_0 \approx 1.0\ bar$, the Mach number of the incident shock-wave is $M_i = 1.3$, the packed-bed length equals $0.14m$, see also Gelfand *et al.* 1986 and Britan *et al.* 1997). This clearly indicates the inapplicability of the "base distance" method for measurements of the wave-front velocity within packed beds of small particle sizes.

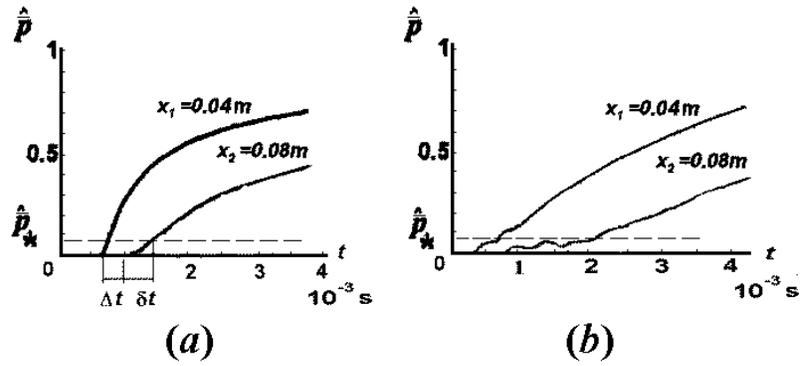

**Figure 4.** Normalized pressure $\hat{\bar{p}}$ vs time $t$ for 2 locations of pressure sensors, $\varphi = 0.4$, $\Lambda_p = 0.25 \cdot 10^{-3}\,m$.

(a) Present modeling, (b) Experiments (Ben-Dor *et al.* 1997).

$$\hat{\bar{p}} = \bar{p}/\Delta p, \hat{\bar{p}}_* = \bar{p}_*/\Delta p,\ \Delta t = (x_2 - x_1)/c_\infty;\ x_2 - x_1\ \text{is the "base distance";}$$

$\Delta p = p_1 - p_0$ is the pressure head through the packed bed; $\Delta t$ is the time duration of the wave front passage of the "base distance", $\delta t$ is the apparent increase in the time duration, $p_1 \approx 2.4\ bar$, $p_0 \approx 1.0\ bar$.

In the present study the following main qualifications have been adopted: (i) the packed bed is a rigid formed material; (ii) the linear approximation is valid; (iii) the wavelength is much larger compared with the grain size; (iv) the high-frequency limit is applicable. Restriction (i) may be satisfied by the choice of the air-filled fixed-bed of solid particles in the experimental set-up. Linear approximation (ii) is valid at small local Mach numbers. Hence, one has using Eq. (13):

$$M = \frac{\bar{u}}{c_\infty} = \frac{\Delta p}{p_0} H(c_\infty t - x) E(\theta), \ \ E(\theta) = \mathrm{erfc}(\theta), \ \theta(x,t) = \frac{x}{2\Lambda} \sqrt{\frac{\nu}{c_\infty} \frac{1}{tc_\infty - x}} \ , \tag{14}$$

where the decrement factor $E < 1$ at $x > 0$ rapidly and monotonically drops to zero with rising $\theta$. Thus, smallness of the dimensionless pressure head, $\Delta p / p_0$, is sufficient for smallness of the local Mach number. For the experimental conditions corresponding to Fig. 4 this condition is violated ($\Delta p / p_0 \approx 1.4$). However, according to Eqs. (14) smallness of the local Mach number also takes place for any fixed location $0 < x < tc_\infty$ at least during sufficiently small time after the arrival of the wave front the point $x$ (i.e. at the early stage of the transient flow). Applicability of restriction (iii) may be justified by comparison with the experimental conditions. Based on Fig. 4, a typical time scale for the waveforms investigated is $\Delta t \sim 0.5 \cdot 10^{-3} s$. This corresponds to wavelength of the order of $\lambda \sim c_\infty \Delta t \approx 0.15 m$, much greater than micro scale pore-size $\Lambda' \sim \Lambda \sim 10^{-3} m$. Concerning restriction (iv), note that the viscous and thermal skin depths, $\delta$ and $\delta'$ ($\delta \sim \delta'$), are assumed to be sufficiently small compared to the corresponding typical pore sizes, $\Lambda$ and $\Lambda'$, respectively ($\Lambda \sim \Lambda'$). This is a quite realistic limit in the high-frequency approximation: $\delta \sim \sqrt{\nu / \omega}$ and $\delta' \sim \sqrt{\nu' / \omega}$. Indeed, for $\Lambda' \sim \Lambda \sim 10^{-3} m$, and for the kinematic viscosity and thermal diffusivity of the air $\nu' \sim \nu \sim 1.5 \cdot 10^{-5} m^2 / s$, we have $\delta / \Lambda \sim \delta' / \Lambda' << 1$ for $\omega / (2\pi) >> \nu / (2\pi \Lambda^2) \approx 2.5 Hz$, or, in terms of characteristic time scales for $t << 1/\omega \sim 0.1 s$. This condition is well satisfied for the typical time scale in the experiments (Fig.4).

**3. Summary and discussion**

The transient regime in shock-induced filtration within packed beds is studied based on the longwave high-frequency approximation for hydraulic resistance force and the effective thermal resistance produced by the viscous and thermal boundary layers near the pore-capillary walls, respectively. The model is capable to describe qualitatively the transient regime flow. Simulated pressure profiles demonstrate good qualitative agreement with the available experimental data (Fig. 4) despite influence of the complex real conditions for the filtrating wave at the input edge of the packed-bed sample, the wave reflection from the back wall of the experimental chamber, measurement errors etc. Furthermore, the strong distortion variation of the modeled pressure wave profile with variation in particle sizes is in qualitative agreement with the measurement results. Thus, the modelling yields that for large particles the initial step-wise pressure profile is slightly changed during the wave propagation, in a manner similar to that in the wide tubes, for small particles the initial pressure profile is strongly distorted with time by the viscous spreading and attenuation (Fig. 3). This strong distortion provides a qualitative explanation for an apparent dependence of the wave-propagation velocity on the particle sizes obtained by the "base distance" method of measurements. Since the wave-propagation velocity equals the gas-sound velocity, $c_\infty$, regardless of the particle size, this illustrates the inapplicability of the "base distance" method widely used for measurements of the wave propagation velocity.

The present study is largely motivated by the gas-dynamic aspects of filtration combustion. Thus, multiplicity of the filtration combustion regimes (the fast combustion in the turbulence regime and the slow combustion in the laminar regime, Babkin 1993) that has not yet found a commonly accepted explanation may be attributed to the instability of unsteady boundary layers developing at the pore capillary walls behind a moving wave inside packed beds (Fig.2). The transient resistance force in terms of the thin unsteady boundary layers developing near the pore capillary may be estimated in the same expressions as in the high-frequency approximation. In neglect of the effective thermal resistance, the thin unsteady boundary layers may be approximated by the classical self-similar solution for the flow of an incompressible fluid near a planar wall starting to move with a constant velocity $u_\infty \equiv \bar{u}(x,t) = const$ from the rest state. This approximation is valid in a far zone behind the wave front, where the solution of problem (1)-(2) is $\bar{p}(x,t) = const$, $\bar{u}(x,t) = const$ and the boundary layer thickness depends only on time, $\delta(x,t) \equiv \delta_\infty(t)$. The approximate boundary-layer solution is as follows: $\tilde{u}(x,y,t) = u_\infty erf\{0.5y/\delta_\infty(t)\}$, $\delta_\infty(t) = \sqrt{\nu t}$ (Landau and Lifshitz 1987, Schlichting 1960). It is known that such unsteady boundary layers lose their stability when the Reynolds number is larger than the critical value $Re_\delta^{(cr)} = \delta^{(cr)} u^* / \nu \approx 487.5$, $\delta^{(cr)}$ is the standard momentum thicknesses, $\delta^{(cr)} \sim \delta_\infty(t)$ (Shtemler 1981). At given characteristic velocity (determined by Eqs. (5)) and fixed viscosity, the value of $Re_\delta^{(cr)}$ determines the critical thickness of the boundary layer, $\delta^{(cr)}$, which, being extrapolated up to the center of the packed bed capillary, may be compared by the order of the values with the equivalent hydraulic radius of the capillary $\Lambda$. The criterion of the micro-level transient instability qualitatively distinguishes two regimes of transition in narrow and wide capillaries. For narrow capillaries ($\Lambda < \delta^{(cr)}$), the unsteady boundary layers have not enough time to lose their stability, and they reach the capillary center due to stable development. For wide capillaries ($\Lambda > \delta^{(cr)}$), the unsteady boundary layers reach the critical value of the boundary-layer thickness when they have lost stability. In the latter case the transient instability makes conventional stability studies of steady-state flows inadequate because filtration flows may not have enough time to reach the steady state. Although such consideration assumes that the boundary-layer thickness is of the order of the pore size, i.e. generally speaking outside of the applicability of the thin boundary-layer approximation, it properly qualitatively reflects the physical mechanism of the transient flow instability.

*Acknowledgements*

The financial support of the Israel Science Foundation under grant 278/03 and United State–Israel Binational Science Foundation under grant 1999248 is gratefully acknowledged. The authors gratefully acknowledge the comments and suggestions made by Denis Lafarge that have significantly improved the content of the current paper.

**Appendix A. Basic relations for the transient regime flow in the high frequency approximation**

Let us consider first the basic equations in the frequencey domain and then in the temporal domain. In the terms of dynamic tortuosity, $\alpha(\omega)$, (Johnson *et al*. 1987) and dynamic compressibility, $\beta(\omega)$, (Champoux and Allard 1991, Lafarge *et al*. 1997). Assuming that the linear approximation is valid, the general governing equations in the frequency domain are as follows:

$$\rho_0 \alpha(\omega) i\omega \, \bar{u} = -\frac{\partial \bar{p}}{\partial x}, \qquad \frac{1}{\gamma p_0}\beta(\omega) i\omega \, \bar{p} = -\frac{\partial \bar{u}}{\partial x}, \qquad (A.1)$$

where $\bar{u}$ and $\bar{p}$ are the macroscopic values of the perturbations of the longitudinal velocity and pressure obtained by averaging corresponding microscopic values near the rest state; $\rho_0$ and $p_0$ are the unperturbed values of the fluid density and pressure at the rest. In a wide range of micro-geometries $\alpha(\omega)$ and $\beta(\omega)$ are the known general model functions given in terms of a few geometrical parameters (Johnson *et al*. 1987, Lafarge *et al*. 1997). In the high frequency limit, when the boundary-layer thicknesses are much less than the pore size there are following asymptotic relations for $\alpha(\omega)$ and $\beta(\omega)$ (Johnson *et al*. 1987, Champoux and Allard 1991):

$$\alpha(\omega) = \alpha_\infty \left\{ 1 + \frac{2}{\Lambda}\sqrt{\frac{\nu}{i\omega}} + O(\frac{\nu}{i\omega}) \right\}, \; \beta(\omega) = 1 + \frac{2(\gamma-1)}{\Lambda'}\sqrt{\frac{\nu'}{i\omega}} + O(\frac{\nu'}{i\omega}), \qquad (A.2)$$

where $\alpha_\infty = 1.37$ is the turtuosity, $\Lambda$ and $\Lambda'$ are the viscous and thermal characteristic lengths, $\nu$ is the kinematic viscosity and $\nu'$ is the thermal diffusivity. In the high-frequency limit, the viscous and thermal skin depths $\delta \sim \sqrt{\nu/\omega}$ and $\delta' \sim \sqrt{\nu'/\omega}$ are assumed sufficiently small compared to typical pore sizes $\Lambda$ and $\Lambda'$. Neglecting the terms of the order of $O(\nu/(\Lambda^2\omega))$ and $O(\nu'/(\Lambda'^2\omega))$ the governing equations (A.2) may be rewritten as follows:

$$\frac{1}{\gamma p_0} i\omega \left(1 + \frac{2(\gamma-1)}{\Lambda'} \sqrt{\frac{\nu'}{i\omega}}\right) \bar{p} = -\frac{\partial \bar{u}}{\partial x}, \quad \rho_0 \alpha_\infty i\omega \, \bar{u} = -\left(1 - \frac{2}{\Lambda} \sqrt{\frac{\nu}{i\omega}}\right) \frac{\partial \bar{p}}{\partial x}. \tag{A.3}$$

Replacing $i\omega$ onto $\partial/\partial t$ yields in the temporal domain the following natural form for the basic system (A.3) together with the linearized mass balance equation:

$$\frac{\partial \bar{\rho}}{\partial t} + \rho_0 \frac{\partial \bar{u}}{\partial x} = 0, \qquad \rho_0 \alpha_\infty \frac{\partial \bar{u}}{\partial t} + \frac{\partial \bar{p}}{\partial x} = F, \quad \bar{p} + G = c_0^2 \bar{\rho} \quad (c_0 = \sqrt{\gamma p_0 / \rho_0}\,), \tag{A.4}$$

where $\bar{\rho}$ is the macroscopic value of the density perturbation; $F$ and $G$ are the hydraulic resistance force and the effective thermal resistance, produced by the viscous and thermal boundary layers, respectively:

$$F = \frac{2\sqrt{\nu}}{\Lambda} \frac{\partial^{1/2} \bar{p}}{\partial x \partial t^{-1/2}}, \qquad G = \frac{2(\gamma-1)\sqrt{\nu'}}{\Lambda'} \frac{\partial^{-1/2} \bar{p}}{\partial t^{-1/2}}. \tag{A.5}$$

Here the fractional derivative is as follows (Samko *et al.* 1993):

$$\frac{\partial^{-1/2} \bar{p}}{\partial t^{-1/2}} \equiv \frac{1}{\sqrt{\pi}} \int_0^t \frac{\bar{p}(x,\theta)}{\sqrt{t-\theta}} \, d\theta. \tag{A.6}$$